\pdfoutput=1
\PassOptionsToPackage{unicode}{hyperref}
\PassOptionsToPackage{hyphens}{url}
\PassOptionsToPackage{dvipsnames,svgnames,x11names}{xcolor}
\documentclass[
]{article}
\usepackage{amsmath,amssymb}
\usepackage{lmodern}
\usepackage{iftex}
\ifPDFTeX
  \usepackage[T1]{fontenc}
  \usepackage[utf8]{inputenc}
  \usepackage{textcomp} 
\else 
  \usepackage{unicode-math}
  \defaultfontfeatures{Scale=MatchLowercase}
  \defaultfontfeatures[\rmfamily]{Ligatures=TeX,Scale=1}
\fi
\IfFileExists{upquote.sty}{\usepackage{upquote}}{}
\IfFileExists{microtype.sty}{
  \usepackage[]{microtype}
  \UseMicrotypeSet[protrusion]{basicmath} 
}{}
\makeatletter
\@ifundefined{KOMAClassName}{
  \IfFileExists{parskip.sty}{%
    \usepackage{parskip}
  }{
    \setlength{\parindent}{0pt}
    \setlength{\parskip}{6pt plus 2pt minus 1pt}}
}{
  \KOMAoptions{parskip=half}}
\makeatother
\usepackage{xcolor}
\usepackage{color}
\usepackage{fancyvrb}

\DefineVerbatimEnvironment{Highlighting}{Verbatim}{commandchars=\\\{\}}
\newenvironment{Shaded}{}{}

\newcommand{\FloatTok}[1]{\textcolor[rgb]{0.25,0.63,0.44}{#1}}

\newcommand{\ImportTok}[1]{\textcolor[rgb]{0.00,0.50,0.00}{\textbf{#1}}}

\newcommand{\KeywordTok}[1]{\textcolor[rgb]{0.00,0.44,0.13}{\textbf{#1}}}
\newcommand{\NormalTok}[1]{#1}
\newcommand{\OperatorTok}[1]{\textcolor[rgb]{0.40,0.40,0.40}{#1}}

\newcommand{\VariableTok}[1]{\textcolor[rgb]{0.10,0.09,0.49}{#1}}

\usepackage{longtable,booktabs,array}
\usepackage{calc} 
\usepackage{etoolbox}
\makeatletter
\patchcmd\longtable{\par}{\if@noskipsec\mbox{}\fi\par}{}{}
\makeatother
\IfFileExists{footnotehyper.sty}{\usepackage{footnotehyper}}{\usepackage{footnote}}
\makesavenoteenv{longtable}
\newlength{\cslhangindent}
\newlength{\csllabelwidth}
\newlength{\cslentryspacingunit} 
\newenvironment{CSLReferences}[2] 
 {
  \setlength{\parindent}{0pt}
  \ifodd #1
  \let\oldpar\par
  \def\par{\hangindent=\cslhangindent\oldpar}
  \fi
  \setlength{\parskip}{#2\cslentryspacingunit}
 }%
 {}
\usepackage{calc}

\ifLuaTeX
\usepackage[bidi=basic]{babel}
\else
\usepackage[bidi=default]{babel}
\fi
\babelprovide[main,import]{american}

\def\languageshorthands#1{}
\ifLuaTeX
  \usepackage{selnolig}  
\fi
\IfFileExists{bookmark.sty}{\usepackage{bookmark}}{\usepackage{hyperref}}
\IfFileExists{xurl.sty}{\usepackage{xurl}}{} 
\hypersetup{
  pdftitle={OpenSkill: A faster asymmetric multi-team, multiplayer
rating system},
  pdfauthor={Vivek Joshy},
  pdflang={en-US},
  colorlinks=true,
  linkcolor={Maroon},
  filecolor={Maroon},
  citecolor={Blue},
  urlcolor={Blue},
  pdfcreator={LaTeX via pandoc}}

\title{OpenSkill: A faster asymmetric multi-team, multiplayer rating
system}

\usepackage[affil-it]{authblk}
\usepackage{orcidlink}
\author[1%
  ]{Vivek Joshy%
    \,\orcidlink{0000-0003-2443-8827}\,%
    }

\affil[1]{Independent Researcher, India}
\date{21 July 2023}

\begin{document}
\maketitle

\hypertarget{summary}{%
\section{Summary}\label{summary}}

In the realm of online gaming, player performance is often measured and
compared through a system called online ranking. This system assigns a
``rank'' to players based on their performance and outcomes in games. A
rank is a distinct level or position within a hierarchy that ostensibly
represents a player's skill relative to others. One common criticism of
this system is the phenomenon known as ``Elo Hell'', a situation where
players find themselves trapped at a certain rank, unable to progress
due to perceived flaws in how the ranking is calculated or due to the
influence of team dynamics (\protect\hyperlink{ref-Elo}{Aeschbach et
al., 2023}).

In the context of player ranking systems, several well-established
models such as Elo and Glicko 2 have set the standard for performance
measurement. However, a limitation becomes evident when these commonly
used systems are considered for games that extend beyond two-player
matchups. The Elo rating system
(\protect\hyperlink{ref-elo_rating_1978}{Elo, 1978}), for one, was
originally formulated for chess, a head-to-head game, and it struggles
to adapt to the complex dynamics of multiplayer scenarios. Similarly,
Glicko 2 (\protect\hyperlink{ref-Glicko2}{Glickman, 2001}), while
offering improvements in rating accuracy and accounting for player
rating volatility still falls short when tasked with accurately
representing the individual contributions within a team-oriented game
setting.

This library represents a concrete implementation of an improved ranking
system, specifically engineered to address the distinctive challenges of
multiplayer gaming environments. It systematically corrects for the
deficiencies of traditional ranking systems, providing a robust solution
that ensures a fair and dynamic evaluation of each player's skill level.
In addition to delivering accuracy on par with implementations of
proprietary models like TrueSkill
(\protect\hyperlink{ref-herbrich2006trueskill}{Herbrich \& Graepel,
2006}), this system distinguishes itself through its enhanced speed in
computing the ranks, facilitating a more responsive and gratifying
player experience.

\hypertarget{statement-of-need}{%
\section{Statement of need}\label{statement-of-need}}

Bayesian inference of skill ratings from game outcomes is a crucial
aspect of online video game development and research. This is usually
challenging because the players' performance changes over time and also
varies based on who they are competing against. Our project primarily
targets game developers and researchers interested in ranking players
fairly and accurately. Nevertheless, the problem that the software
solves applies to any context where you have multiple players or
entities and you need to track their skills over time while they compete
against each other.

The OpenSkill library furnishes a versatile suite of models and
algorithms designed to support a broad spectrum of applications. While
popular use cases include assisting video game developers and
researchers dealing with multi-agent scripting environments like Neural
MMO (\protect\hyperlink{ref-suarez2019neural}{Suarez et al., 2019}), its
practical use extends far beyond this particular domain. For instance,
it finds substantial utilization in recommendation systems, where it
efficiently gauges unique user behaviours and preferences to suggest
personalized recommendations. The matchmaking mechanisms in ranking of
sports players as seen by Opta Analyst
(\protect\hyperlink{ref-Rico_2022}{Rico, 2022}) and dating apps are
another area where OpenSkill proves crucial, ensuring an optimal pairing
based on the comparative ranking of user profiles' competencies.

Derived from the research paper by Weng and Lin
(\protect\hyperlink{ref-JMLR:v12:weng11a}{Weng \& Lin, 2011}), OpenSkill
offers a pure Python implementation of their Bayesian approximation
method for probabilistic models of ranked data. OpenSkill attempts to
solve the same problems TrueSkill does. TrueSkill however employs factor
graphs to model the probability distributions of players' skills,
updating their ranks through Bayesian inference after each game by
evaluating the likelihood of observed outcomes.

Similar to TrueSkill this library is specifically designed for
asymmetric multi-faction multiplayer games. In the games it's intended
for, the term ``asymmetric'' means that teams might have varying numbers
of players. For example, one team could have three players while another
has just one. This creates an uneven playing field where the challenge
is to balance these differences. The term ``multi-faction'' means that
there are several distinct teams or groups within a single game. Unlike
simple one-on-one contests, these games feature multiple teams, each
potentially with a different number of players, all competing in the
same match. This library aims to assess and balance player skill in such
dynamic and complex game environments.

OpenSkill boasts several advantages over implementations of proprietary
models like TrueSkill. Notably, it delivers faster rating updates, with
3 times the performance of the popular Python open-source implementation
of TrueSkill as seen in Lee (\protect\hyperlink{ref-Lee_2018}{2018}).
OpenSkill also includes five distinct models, each with its unique
characteristics and tradeoffs. While all the models are general purpose,
the recommended model for most use cases is Plackett-Luce. This model
extends the regular Plackett-Luce as described in Guiver \& Snelson
(\protect\hyperlink{ref-Plackett_Luce}{2009}) by incorporating variance
parameters to account for the probability that a certain team is the
winner among a set of competing teams.

The Plackett-Luce model can be thought of as a generalized extension of
the Braldey-Terry model originally introduced in Bradley \& Terry
(\protect\hyperlink{ref-Bradley_Terry}{1952}). Both models follow
logistic distribution, while in contrast, the Thurstone-Mosteller model
follows the Gaussian distribution. Both models can be also used with
partial pairing and full pairing approaches for rating updates. Partial
pairing models engage only a subset of players who are paired with each
other during rating updates. This strategy considerably improves
computational efficiency while sacrificing a certain level of accuracy.
On the other hand, full pairing models leverage all available
information within the paired data to make precise rating updates at the
cost of increased computational requirements.

\hypertarget{usage}{%
\section{Usage}\label{usage}}

To install the library simply \texttt{pip\ install\ openskill} and
import the library. A conventional example of usage is given below:

\begin{Shaded}
\begin{Highlighting}[]

\OperatorTok{\textgreater{}\textgreater{}\textgreater{}} \ImportTok{from}\NormalTok{ openskill.models }\ImportTok{import}\NormalTok{ PlackettLuce}
\OperatorTok{\textgreater{}\textgreater{}\textgreater{}}\NormalTok{ model }\OperatorTok{=}\NormalTok{ PlackettLuce()}
\OperatorTok{\textgreater{}\textgreater{}\textgreater{}}\NormalTok{ model.rating()}
\NormalTok{PlackettLuceRating(mu}\OperatorTok{=}\FloatTok{25.0}\NormalTok{, sigma}\OperatorTok{=}\FloatTok{8.333333333333334}\NormalTok{)}
\OperatorTok{\textgreater{}\textgreater{}\textgreater{}}\NormalTok{ r }\OperatorTok{=}\NormalTok{ model.rating}
\OperatorTok{\textgreater{}\textgreater{}\textgreater{}}\NormalTok{ [[a, b], [x, y]] }\OperatorTok{=}\NormalTok{ [[r(), r()], [r(), r()]]}
\OperatorTok{\textgreater{}\textgreater{}\textgreater{}}\NormalTok{ [[a, b], [x, y]] }\OperatorTok{=}\NormalTok{ model.rate([[a, b], [x, y]])}
\OperatorTok{\textgreater{}\textgreater{}\textgreater{}}\NormalTok{ a}
\NormalTok{PlackettLuceRating(mu}\OperatorTok{=}\FloatTok{26.964294621803063}\NormalTok{, sigma}\OperatorTok{=}\FloatTok{8.177962604389991}\NormalTok{)}
\OperatorTok{\textgreater{}\textgreater{}\textgreater{}}\NormalTok{ x}
\NormalTok{PlackettLuceRating(mu}\OperatorTok{=}\FloatTok{23.035705378196937}\NormalTok{, sigma}\OperatorTok{=}\FloatTok{8.177962604389991}\NormalTok{)}
\OperatorTok{\textgreater{}\textgreater{}\textgreater{}}\NormalTok{ (a }\OperatorTok{==}\NormalTok{ b) }\KeywordTok{and}\NormalTok{ (x }\OperatorTok{==}\NormalTok{ y)}
\VariableTok{True}
\end{Highlighting}
\end{Shaded}

Each player has a \texttt{mu} and a \texttt{sigma} value corresponding
to their skill (\(\mu\)) and uncertainty (\(\sigma\)) in skill.
Comparisons between two players can be done by calling the
\texttt{ordinal()} method. In this case it would be on the instances of
\texttt{PlackettLuceRating}.

\hypertarget{benchmarks}{%
\section{Benchmarks}\label{benchmarks}}

A reproducible set of benchmarks is available in the \texttt{benchmark/}
folder at the root of the openskill.py repository. Simply run the
appropriate Jupyter Notebook file to run the relevant benchmark.

Using a dataset of Overwatch
(\protect\hyperlink{ref-joshy_2023_overwatch}{Joshy, 2023}) matches and
player info, OpenSkill predicts the winners competitively with
TrueSkill.

For games restricted to at least 2 matches per player:

\begin{longtable}[]{@{}ccc@{}}
\toprule\noalign{}
& OpenSkill - PlackettLuce & TrueSkill \\
\midrule\noalign{}
\endhead
\bottomrule\noalign{}
\endlastfoot
Correct Matches & 556 & 587 \\
Incorrect Matches & 79 & 48 \\
Accuracy & \textbf{87.56\%} & \textbf{92.44\%} \\
Runtime Duration & \textbf{0.97s} & 3.41s \\
\end{longtable}

When restricted to 1 match per player:

\begin{longtable}[]{@{}ccc@{}}
\toprule\noalign{}
& OpenSkill - PlackettLuce & TrueSkill \\
\midrule\noalign{}
\endhead
\bottomrule\noalign{}
\endlastfoot
Correct Matches & 799 & 830 \\
Incorrect Matches & 334 & 303 \\
Accuracy & \textbf{70.52\%} & \textbf{73.26\%} \\
Runtime Duration & \textbf{17.64s} & 58.35 \\
\end{longtable}

Using a dataset of chess matches, we also see a similar trend, where
OpenSkill gives a similar predictive performance to TrueSkill, but in
less time.

It should be noted that the difference in speed may be partially due to
the the efficiency of the TrueSkill implementation in question. For
instance, switching to Scipy backend in the TrueSkill implementation
slows the inference to around 8 seconds even though we should be
expecting a speedup since Scipy drops into faster C code.

Finally, running the project against a large dataset of PUBG online
matches results in a Rank-Biased Overlap
(\protect\hyperlink{ref-10.1145ux2f1852102.1852106}{Webber et al.,
2010}) of \textbf{64.11} and an accuracy of \textbf{92.03\%}.

\hypertarget{discussion}{%
\section{Discussion}\label{discussion}}

Our OpenSkill library has demonstrated significant improvements over
proprietary models in terms of both speed and efficiency. However, we
recognize that there are still areas that warrant further exploration
and improvement.

One such area is partial play. Ideally, a comprehensive skill ranking
system should take into account both the winning and losing side of a
game and adjust their ratings accordingly. Partial play, where only a
subset of players are engaged during a match, presents a unique
challenge in this regard. While it is theoretically easy to implement
this feature, the lack of relevant data makes it difficult for us to
verify its efficacy. Consequently, any modifications we make to such
models run the risk of overfitting the available data. The absence of a
clearly defined metric for partial play further complicates matters, as
different groups interpret it in various ways. Our interpretation of
partial play pertains to the duration a player participates in a game,
but significant work is required to operationalize this concept in a
tangible way within our library.

More substantially, as of now, OpenSkill does not support weight
integration, where weights represent a player's contributions to an
overall victory. The ability to assign different significance to
different players based on their contributions could greatly improve the
accuracy of a player's resulting skill rating. We realize the value of
this feature, and it is a primary area of focus in our ongoing
improvements to the library.

On a positive note, OpenSkill does indeed support time decay, an
important aspect of maintaining an accurate skill rating system. Over
time, a player's skill can decrease due to inactivity; our library
allows users to adjust the sigma value accordingly. This feature ensures
that our library maintains its adaptability and relevance even when
faced with variable player engagement levels.

Despite these limitations, our OpenSkill library remains a powerful tool
for video game developers and researchers tasked with competently
evaluating player skills. It addresses several long-standing issues
encountered in multiplayer video game ranking systems. By continuously
seeking out improvements and refining our approach, we hope to make
OpenSkill an ever more effective and flexible resource in the world of
online gaming.

\hypertarget{related-packages}{%
\section{Related Packages}\label{related-packages}}

This project was originally a direct port of the openskill.js project
(\protect\hyperlink{ref-Busby_2023}{Busby, 2023}) from Javascript to
Python. However, we have deviated slightly from their implementation in
that we focus more on Python-specific features, and thorough
documentation of every object. All documented objects have the
mathematical formulas from their respective papers included for easier
inspection of code. We also provide an easy way to customize all the
constants used in any model very easily. There are also published ports
of OpenSkill in Elixir
(\protect\hyperlink{ref-PhilihpOpenSkillElixir}{Busby, 2020}), Kotlin
(\protect\hyperlink{ref-BrezinaOpenSkillKotlin}{Brezina, 2022}) and Lua
(\protect\hyperlink{ref-BstummerOpenSkillLua}{\emph{{G}it{H}ub -
Bstummer/Openskill.lua --- Github.com}, 2022}) on GitHub.

When comparing our OpenSkill to similar packages like that of Lee's
TrueSkill implementation, we also provide support for PyPy 3, which uses
a Just-In-Time compiler as opposed to the standard CPython
implementation. We also support strict typing of objects, to enable
auto-completion in your Integrated Development Environments (IDEs). Our
development workflow also requires a test coverage of 100\% for any code
to be merged. This serves as a starting point to prevent erroneous math
from making it into the library.

\hypertarget{acknowledgements}{%
\section{Acknowledgements}\label{acknowledgements}}

We extend our sincere gratitude to Philihp Busby and the openskill.js
project for their valuable contributions without which this project
would not have been possible. Additionally, their inputs and
contributions to the prediction methods, have significantly enhanced its
speed and efficiency. Special acknowledgment also goes to Jas Laferriere
for their critical contribution of the additive dynamics factor. Your
collective efforts have been instrumental in improving our work.

\hypertarget{references}{%
\section*{References}\label{references}}
\addcontentsline{toc}{section}{References}

\hypertarget{refs}{}
\begin{CSLReferences}{1}{0}
\leavevmode\vadjust pre{\hypertarget{ref-Elo}{}}%
Aeschbach, L. F., Kayser, D., Berbert De Castro Hüsler, A., Opwis, K.,
\& Brühlmann, F. (2023). The psychology of esports players' ELO hell:
Motivated bias in league of legends and its impact on players'
overestimation of skill. \emph{Computers in Human Behavior}, \emph{147},
107828. \url{https://doi.org/10.1016/j.chb.2023.107828}

\leavevmode\vadjust pre{\hypertarget{ref-Bradley_Terry}{}}%
Bradley, R. A., \& Terry, M. E. (1952). {Rank Analysis of incomplete
block designs: {T}he method of paired comparisons}. \emph{Biometrika},
\emph{39}(3-4), 324--345.
\url{https://doi.org/10.1093/biomet/39.3-4.324}

\leavevmode\vadjust pre{\hypertarget{ref-BrezinaOpenSkillKotlin}{}}%
Brezina, J. (2022). \emph{{G}it{H}ub - brezinajn/openskill.kt ---
github.com}. \url{https://github.com/brezinajn/openskill.kt}.

\leavevmode\vadjust pre{\hypertarget{ref-PhilihpOpenSkillElixir}{}}%
Busby, P. (2020). \emph{{G}it{H}ub - philihp/openskill.ex: {E}lixir
implementation of {W}eng-{L}in {B}ayesian ranking, a better,
license-free alternative to {T}rue{S}kill --- github.com}.
\url{https://github.com/philihp/openskill.ex}.

\leavevmode\vadjust pre{\hypertarget{ref-Busby_2023}{}}%
Busby, P. (2023). A faster, open-license alternative to microsoft
TrueSkill. In \emph{GitHub}.
\url{https://github.com/philihp/openskill.js}

\leavevmode\vadjust pre{\hypertarget{ref-elo_rating_1978}{}}%
Elo, A. E. (1978). \emph{The rating of chessplayers, past and present}.
Arco Pub. ISBN:~9780668047210

\leavevmode\vadjust pre{\hypertarget{ref-BstummerOpenSkillLua}{}}%
\emph{{G}it{H}ub - bstummer/openskill.lua --- github.com}. (2022).
\url{https://github.com/bstummer/openskill.lua}.

\leavevmode\vadjust pre{\hypertarget{ref-Glicko2}{}}%
Glickman, M. E. (2001). Dynamic paired comparison models with stochastic
variances. \emph{Journal of Applied Statistics}, \emph{28}(6), 673--689.
\url{https://doi.org/10.1080/02664760120059219}

\leavevmode\vadjust pre{\hypertarget{ref-Plackett_Luce}{}}%
Guiver, J., \& Snelson, E. (2009). Bayesian inference for plackett-luce
ranking models. \emph{Proceedings of the 26th Annual International
Conference on Machine Learning}, 377--384.
\url{https://doi.org/10.1145/1553374.1553423}

\leavevmode\vadjust pre{\hypertarget{ref-herbrich2006trueskill}{}}%
Herbrich, R., \& Graepel, T. (2006). \emph{TrueSkill(TM): A {B}ayesian
skill rating system} (MSR-TR-2006-80).
\url{https://www.microsoft.com/en-us/research/publication/trueskilltm-a-bayesian-skill-rating-system-2/}

\leavevmode\vadjust pre{\hypertarget{ref-joshy_2023_overwatch}{}}%
Joshy, V. (2023). \emph{OverWatch match data} (Version 1.0.0) {[}Data
set{]}. Zenodo. \url{https://doi.org/10.5281/zenodo.10359600}

\leavevmode\vadjust pre{\hypertarget{ref-Lee_2018}{}}%
Lee, H. (2018). An implementation of the TrueSkill rating system for
python. In \emph{sublee/trueskill: An implementation of the TrueSkill
rating system for Python}. \url{https://github.com/sublee/trueskill}

\leavevmode\vadjust pre{\hypertarget{ref-Rico_2022}{}}%
Rico, Y. G. (2022). Her ranking is a 10 but her skill rating says... In
\emph{The Analyst}. Opta Analyst.
\url{https://theanalyst.com/eu/2022/08/true-tennis-skill-ratings/}

\leavevmode\vadjust pre{\hypertarget{ref-suarez2019neural}{}}%
Suarez, J., Du, Y., Isola, P., \& Mordatch, I. (2019). \emph{Neural MMO:
A massively multiagent game environment for training and evaluating
intelligent agents}. \url{https://arxiv.org/abs/1903.00784}

\leavevmode\vadjust pre{\hypertarget{ref-10.1145ux2f1852102.1852106}{}}%
Webber, W., Moffat, A., \& Zobel, J. (2010). A similarity measure for
indefinite rankings. \emph{ACM Trans. Inf. Syst.}, \emph{28}(4).
\url{https://doi.org/10.1145/1852102.1852106}

\leavevmode\vadjust pre{\hypertarget{ref-JMLR:v12:weng11a}{}}%
Weng, R. C., \& Lin, C.-J. (2011). A {B}ayesian approximation method for
online ranking. \emph{Journal of Machine Learning Research},
\emph{12}(9), 267--300. \url{http://jmlr.org/papers/v12/weng11a.html}

\end{CSLReferences}

\end{document}